\documentclass[prl,twocolumn,showpacs]{revtex4}
\usepackage{amsmath,amssymb}
\usepackage{epsfig}

\begin{document}

\newlength{\mylen}
\setlength{\mylen}{\textwidth}
\addtolength{\mylen}{-1cm}
\newcommand{\bea}{\begin{eqnarray}}
\newcommand{\eea}{\end{eqnarray}}
\newcommand{\spaceint}[2]{\int_{#1} d^3 #2 \;}
\newcommand{\vect}[1]{\mathbf{#1}}
\newcommand{\vat}{V^{\rm att}}
\newcommand{\di}{\displaystyle}
\newcommand{\rp}{r_{||}}
\newcommand{\ep}{\varepsilon_\Pi}
\newcommand{\ef}{\varepsilon_F}
\newcommand{\emix}{\varepsilon_{\rm m}}
\newcommand{\ev}{\varepsilon_{\rm rep}}
\newcommand{\Ezp}{E_{z,0^+}}
\newcommand{\Ezm}{E_{z,0^-}}
\newcommand{\Ep}{E_{||,0}}
\newcommand{\ro}{r_{0,{\rm ref}}}


\title{Charge renormalization for effective interactions of colloids
at water interfaces}

\author{Derek Frydel and S. Dietrich}
\affiliation{Max--Planck--Institut f\"ur Metallforschung, Heisenbergstr.~3,
  D--70569 Stuttgart, Germany,}
\affiliation{Institut f\"ur Theoretische und Angewandte
  Physik, Universit\"at Stuttgart, Pfaffenwaldring 57, D--70569 Stuttgart, Germany}
\author{Martin Oettel}
\affiliation{Institut f\"ur Physik, Johannes-Gutenberg-Universit\"at Mainz,
WA 331, 55099 Mainz, Germany}

\date{\today}

\begin{abstract}
 We analyze theoretically the electrostatic interaction of surface--charged
 colloids at water interfaces with special attention to the experimentally
 relevant case of large charge 
 densities on the colloid--water interface. Whereas linear theory predicts  
 an effective dipole potential the strength of which is proportional to the
 square of the product of 
 charge density and screening length, nonlinear charge renormalization effects
 change this dependence to a weakly logarithmic one. These results appear to
 be particularly  relevant for structure formation at air--water interfaces
 with arbitrarily shaped colloids. 
\end{abstract}


\maketitle

The effective interactions of colloids trapped at fluid interfaces reveal
qualitatively new features when compared to the ones in colloidal bulk
solutions.  First, there is the possibility of long--ranged
capillary attractions mediated by deformations of the interface
\cite{Zen06,Dom05}. Second, many colloids carry a significant amount of charge
(e.g. charge--stabilized polymeric colloids, mineralic disks, proteins)
and the exponentially screened electrostatic
interactions in ionic bulk solvents become longer--ranged at interfaces
between water and a nonpolar medium (typically air or oil).
At such interfaces the colloids exhibit effective
dipole--like repulsions which lead to the stabilization of two--dimensional
crystals even at low surface coverages \cite{Pie80}. These effective dipoles
originate from colloidal surface charges on the water side and a cloud of
screening ions in the water phase which is asymmetric with respect to the
interface plane.  Within a simple model (the only
analytically tractable one) the colloids
are approximated as equal point charges $q$ located in the interface plane and
the water phase is treated as a linearly screening medium.  
To leading order the interaction between two charges $q$ in the
interface plane at separation $d$ is given by \cite{Hur85} 
\bea
 \label{eq:hurd}
 U(d) = q^2 \frac{\epsilon_1}{2\pi\,\epsilon_0\,\epsilon_2^2} 
\frac{\kappa^{-2}}{d^3}\;.
\eea  
Here, $\epsilon_1$ and $\epsilon_2$ are the permittivities of the 
nonpolar medium and water, respectively, and $\epsilon_0$ is the dielectric
constant of vacuum.  According to this linear model the repulsion depends
quadratically  on the Debye screening length 
$\kappa^{-1}=(\epsilon_2\epsilon_0/(2\beta c_0 e^2))^{1/2}$
where $c_0$ is the concentration of monovalent ions in bulk water, $e$ is the
elementary charge, and $\beta^{-1}=k_B T$.  On this basis one would expect the
repulsion $U \propto c_0^{-1}$ to become significantly weaker upon adding
electrolytes.  
Various studies of colloidal aggregation at interfaces have used the
predictions of the linear model for quantitative analysis of
experimental results (see, e.g., Refs. \cite{Ave00,Ave02}).  
In Ref.~\cite{Ave02} $q$ in Eq.~(\ref{eq:hurd}) was replaced by 
$q_{\rm eff}\propto\kappa^{-1}$ to account for geometric effects of a charged
spherical colloid, which leads to $U \propto c_0^{-2}$.

The high colloidal surface charge densities $\sigma_c$ on the water side of
experimentally used colloids (easily up to 0.5 $e$/nm$^2$) invalidate 
the naive use of the linearized Debye--H{\"u}ckel (DH) model with bare
charges.  Strong charge renormalization will occur due to the nonlinear
contributions of the governing Poisson--Boltzmann equation (PB) in the water
phase.  The renormalization procedure (based on the separation of length
scales) consists of the identification of the appropriate corresponding linear
solution of the PB problem at distances $>\kappa^{-1}$ from the charges.
There the electrostatic potential $\Phi$ is small and linear DH electrostatics
holds: $\nabla^2\Phi\simeq\kappa^2\Phi$.  
For a uniformly charged wall or sphere, this solution has the same functional
form as if the entire problem is solved within the
linear theory and the nonlinear effects alter only the prefactor.  This
prefactor leads to a renormalized, effective charge \cite{L02}.
For non-spherical charged bodies the map between the DH solution and the PB
solution in the linear region requires a selection of the appropriate boundary
conditions at the charged object such that the DH and the PB solution match at
the far field \cite{TBA02}.  In the limit $\sigma_c\to\infty$ of the surface
charge density the renormalized DH potential at the colloid surface levels 
off at a constant regardless of the geometry of the charged body.  

The renormalization of charges at an interface is expected to differ from
that in the bulk due to the proximity of a nonpolar phase which induces an
algebraic decay of the electrostatic field near the interface;  to a large
extent its strength is determined by the potential within the screening
length.  In order to study the effect of an interface on the renormalization
we have chosen the experimentally relevant system of a charge--stabilized
colloidal sphere trapped at an interface with water.  The renormalized dipole
field can be described in terms of a single
renormalized parameter given by the effective charge $q_{\rm eff}$.  We find
that the ratio $q_{\rm eff}/q$ factorizes into a geometric part (describable
by a linear theory) which takes into account the geometry of a charged object,
i.e., the charge distribution at the colloid-water interface, and a nonlinear 
part which is described by the analytically solvable case (within PB theory)
of a charged wall, thus being independent of the colloid shape and the contact
angle.  In contrast to the case of colloids in the {\em bulk} we find that 
$q_{\rm eff}$ does not level off for highly charged particles.  Also the
functional dependence of $q_{\rm eff}$ on $\kappa$ differs from the
bulk case;  nonetheless $q_{\rm eff}$ remains an increasing function of 
$\kappa$ \cite{BTA02}.  As a consequence, the effective repulsion given by
Eq.~(\ref{eq:hurd}) becomes only weakly dependent on the screening
length.  

 


\begin{figure}
 \epsfig{file=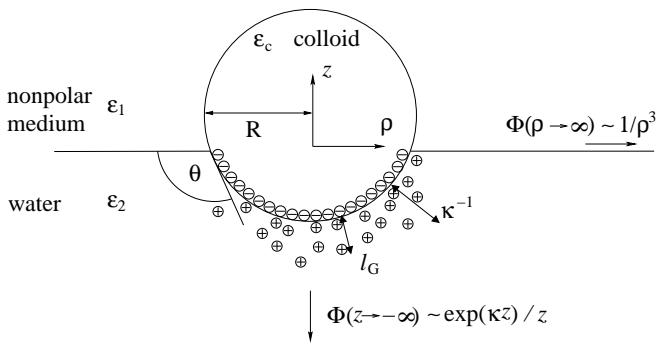, width=\columnwidth}
 \caption{Side view of a single colloid (homogeneously charged on the water
   side) trapped at the interface.  Most of the counterions are confined in a
 layer close to the colloid surface with a width of the order of the
 Gouy--Chapman length $l_{\rm G}=2\epsilon_2\epsilon_0/(\beta e \sigma_c)$. In
 many colloidal experiments, $l_{\rm G} (\approx 1$ nm) 
 $<$ $\kappa^{-1} (\approx 1 \dots 300$ nm) $<$ $R (\approx 1$ $\mu$m).}
 \label{fig:pb_ref}
\end{figure}

{\em The model.} For a single spherical colloid of radius $R$ trapped at an
interface as indicated in Fig.~\ref{fig:pb_ref} we have solved the
electrostatic problem given by the Poisson--Boltzmann equation in the water
phase, ${\nabla^*}^2\phi={\kappa^*}^2\sinh[\phi] $, and the Laplace equation
in the oil phase, ${\nabla^*}^2\phi=0$.  Here, $\phi = e\beta\Phi$, 
$\nabla^*=R\nabla$ and $\kappa^*=\kappa R$ are the dimensionless electrostatic
potential, gradient operator and screening length, respectively.  At the 
water--oil and the colloid--oil interface the tangential electric field and
the normal electric displacement are continuous, while at the colloid--water
interface the normal electric displacement has a jump $\sigma_c$. The 
differential equations with the appropriate boundary conditions are solved
using the finite element method package FEMLAB \cite{FEM}.
In order to determine the potential at large distances from the particle we
have chosen the computational space to be $8000\,R$  so that the boundary
conditions enclosing the box do not influence the data of interest. 
The nominal charge on the colloid is $q=\sigma_c \;2\pi R^2(1+\cos\theta)$. We
have determined the effective charge through equating the asymptotics of the
potential in the water--oil interfacial plane to the asymptotics of the
potential for the point charge in Debye-H{\"u}ckel approximation: 
$q_{\rm eff}=\lim_{\rho\to\infty} (2\pi\epsilon_0\epsilon_2^2/\epsilon_1) 
(\rho^3/\,e\beta\kappa^{-2})\,\phi(\rho,z=0)$. Finite box size effects 
become visible at a distance  $\rho=500\,R$  from the colloid; thus all our
data are taken within this range.  The electrostatic interaction between two
colloids at separation $d$ is indeed given by Eq.~(\ref{eq:hurd}) to leading
order in $d$, with $q$ replaced by $q_{\rm eff}$. This can be shown by a
direct calculation of the force via a pressure tensor integration over the
midplane (symmetry plane) between the two colloids. For fixed permitivities,
the ratio $q_{\rm eff}/q = g(\kappa^*,\sigma_c^*;\theta)$ defines a
renormalization function which depends on $\kappa^*$, the dimensionless charge
density $\sigma_c^*=(e \beta R/(\epsilon_0\epsilon_2))\,\sigma_c$ and
$\theta$. 

\begin{figure}
  \epsfig{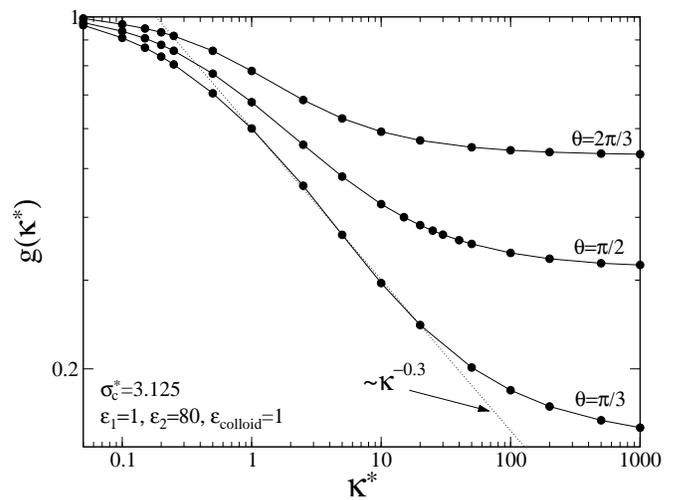}
  \caption{The renormalization function in the linear regime.}
  \label{fig:lin}
\end{figure}

{\em The linear Debye-H{\"u}ckel regime.}
 The linear regime holds if $\phi \ll 1$ everywhere and corresponds
to $\sigma_c^* \kappa^{*-1} \ll 1$.  (The retrieval of the linear regime in
this limit can be confirmed from the exact solution for the charged wall
 model.)  In this regime, the renormalization function is independent of
 $\sigma_c^*$: $g\to g_{\rm lin}(\kappa^*,\theta)$.
The variation of $g_{\rm lin}$ with 
$\kappa^*$ and $\theta$ is moderate and thus the renormalization function
is of the order 1 (see Fig.~\ref{fig:lin}). 
The variation of $g_{\rm lin}$ resembles a weak effective power--law for
a limited range of $\kappa^*$ but it is clearly inconsistent with the
proposal in Ref.~\cite{Ave02} that it should vary $\propto \kappa^{*-1}$ in
the range $1<\kappa^{*-1}<\infty$.  
The weak dependence of $g_{\rm lin}$ on $\kappa^*$  reflects the fact that the
electrostatic field originating from the surface charges ``escapes" to the
insulator phase both through the colloid and, to some extent, through the
electrolyte.  At large $\kappa^*$  the electrolyte ``escape" route is blocked
due to the thick counterion cloud surrounding the charged colloid and so the
dependence of $g_{\rm lin}$ on $\kappa^*$ disappears.  The inadequate
assumption of Ref.~\cite{Ave02} is that the electrolyte ``escape'' route is
the only one except for the field originating from the charges near the
three--phase contact line. 

{\em The nonlinear regime.}
As inferred from the linear regime the geometric contributions to the
effective charge do not have a strong influence on $g$ ($g_{\rm lin}$ is of 
the order of 1 for various contact angles as shown in Fig.~\ref{fig:lin}).  This
encourages us to deduce some general properties of
$g$ without solving the full problem explicitly.  In typical colloidal
experiments \cite{Ave00} the radius of the colloid is of the order of 1 $\mu$m
and thus is much larger than the screening length for
electrolyte concentrations $c_0>10^{-5}$ M ($\kappa^{-1} < 0.1$ $\mu$m).
Therefore close to the colloid surface at the water side the electrostatic
problem is similar to that for a charged wall in electrolyte. 

Since for a charged wall the potential outside the screening length levels off
at large $\sigma_c$ \cite{L02} and the strength of the potential in the
linear regime is $\sigma_c^*\kappa^{*-1}$, this implies that for fixed
$\kappa^{*-1}$ and large 
$\sigma_c^*$, $\lim_{\sigma_c^*\to \infty} g \to 0$ (in order to satisfy
$\sigma_{c,{\rm eff}}^*\kappa^{*-1}={\rm const}$), and that for fixed, large
$\sigma_c^*$, $g$ must increase with $\kappa^{*}$, i.e., $g$ must increase upon
adding electrolyte.  Extrapolating these results for the
charged wall to the present situation we find 
$g\approx 4/(\sigma_c^*\kappa^{*-1})$,
i.e., $q_{\rm eff}$ is proportional to the screening length. Thus 
the interaction potential between two colloids (Eq.~(\ref{eq:hurd}) with 
$q \to q_{\rm eff}$) is {\em independent of the screening length and thus of
the electrolyte concentration}, at least within this crude ``wall
approximation".

\begin{figure}
  \epsfig{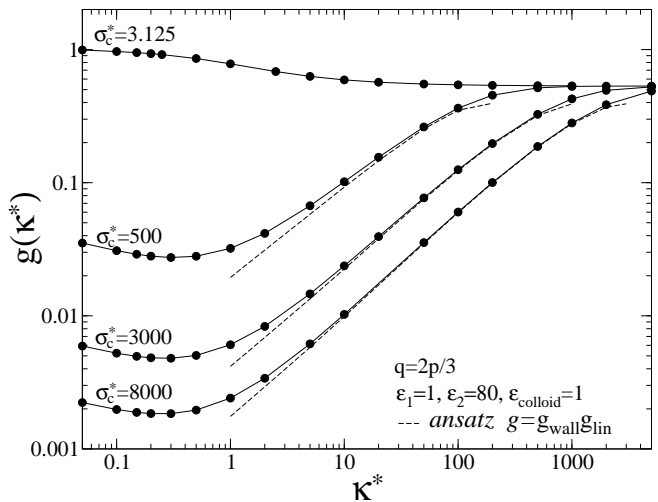}
  \caption{The charge renormalization function in the nonlinear regime.
  For a colloid of radius $R=1$ $\mu$m, the two dimensionless charge densities 
  $\sigma_c^*=500$ and 8000 correspond
  to charge densities of 0.9 and 15 $\mu$C/cm$^2$ which approximately bracket
  the charge densities occurring on polymeric colloids.}
  \label{fig:nonlin}
\end{figure}

Our numerical results show, however, that, unlike in the bulk case, the
effective charge does not level off but increases slowly: 
$q_{\rm eff} \propto\ln\sigma_c^*$. 
This can be understood in terms of a second, somewhat more refined ``wall
approximation".  At the interface, the asymptotic behavior of the potential is
determined by the electric field which ``escapes" to the oil phase. 
The escaping field strength is proportional to the potential right at the
colloid surface on the water side because the escaping field lines originate
there.  Thereby we can approximate the charge renormalization function from
the contact potential at the wall 
$g_{\rm wall}=\sigma^*_{c,{\rm eff}}/\sigma^*_c$. 
The relation between the surface charge and the potential at contact
$\phi^c_{\rm wall}$ for a charged wall is
$\sigma_c^*=2\kappa^*\sinh[\phi^c_{\rm wall}/2]$ \cite{L02}; in the linear 
limit (i.e., small $\sigma_c^*/\kappa^*$) this reduces to
$\phi^c_{\rm wall} = \sigma^*_{c}\kappa^{*-1}$ and in the highly nonlinear
limit (i.e., $\sigma_c^*/\kappa^*$ large) 
$\phi^c_{\rm wall} = 2 \ln (\sigma_c^*\kappa^{*-1})$.   
The effective surface charge is obtained by equating the two limiting cases 
leading to
$g_{\rm wall} = 2(\ln(\sigma_c^*\kappa^{*-1}))/(\sigma_c^*\kappa^{*-1})$.
However, the full renormalization function $g$ contains in addition the
geometric contributions unaccounted for by the wall approximation.  We
augment the nonlinear ``wall" part by the linear ``geometry" part, which 
we have shown in Fig.~\ref{fig:lin}: 
$g\approx g_{\rm wall}(\sigma_c^*,\kappa^{*})\,g_{\rm lin}(\kappa^{*},\theta)$.   
In the strongly nonlinear regime this {\em ansatz} describes our full numerical
data for $g$ rather well (see Fig.~\ref{fig:nonlin}). 
The wall model approximation of the renormalization function can be
corroborated in an
alternative, more involved determination of $\sigma_{c,{\rm eff}}^*$ by
calculating
the effective dipole generated by the surface charges and the counterion cloud.
The latter approach gives rise to corrections $O(\kappa^{*-1})$ which explain the 
behavior of $g$ for small $\kappa^*$.  The failure of the {\em ansatz} for
large $\kappa^*$ reflects the disappearance of the nonlinear effects in this
range.


Inserting $q_{\rm eff}$ (as obtained from the wall model) into 
Eq.~(\ref{eq:hurd}) provides the interaction potential, exhibiting a weak
dependence on the screening length:      
\begin{eqnarray}
 \beta U(d) &\approx& \frac{8\epsilon_1}{\epsilon_2}\frac{R}{\lambda_B}\, 
   \cos^4\left(\frac{\theta}{2}\right) \frac{R^3}{d^3}\, 
   \ln^2\left(\frac{\sigma_c^*}{\kappa^*}\right) \,g_{\rm lin}^2(\kappa^*,\theta). \quad
 \label{eq:u_nonlin}
\end{eqnarray}
Here $\lambda_B=\beta e^2/(4\pi \epsilon_2\epsilon_0) \approx 0.7$ nm is the
Bjerrum length for water.  As discussed before $g_{\rm lin}$ becomes a
constant of the order of 1 for large $\kappa$ and the $\kappa$-dependence of 
$U$ is contained only in the wall term 
$U\propto\ln^2[{\sigma_c^*}{\kappa^{*-1}}]$. 
The comparison with the predictions of the linear theory, 
$U\propto(\sigma_c^{*}\kappa^{*-1})^2$, shows that the nonlinear PB theory
yields a drastically changed dependence on both the charge density and the
screening length. 

{\em Comparison with experiment.} 
There are a number of publications concerning the aggregation and compression
behavior of colloidal mononolayers.  However, direct or indirect measurements
of pair potentials are rather limited. 
Reference \cite{Che06} reports
results for two batches of polystyrene (PS) particles: (a) charged sulfate
groups, $R=0.55$ $\mu$m, $\sigma_c=12.5$ $\mu$C/cm$^2$ and (b) charged carboxyl
groups, $R=0.5$ $\mu$m, $\sigma_c=2.8$ $\mu$C/cm$^2$, both investigated at
an interface between air and ultrapure water ($\kappa^* \approx 1$). 
The tail of the repulsive potential (obtained by inverting pair correlation
functions) was fitted to a dipole form (see Tab.~\ref{tab}).  Comparison with
the present renormalized theory (Eq.~\ref{eq:u_nonlin}) requires knowledge of
$\theta$. Two different visual methods applied to sulfonated PS particles at
the air--water interface \cite{Ave00,Pau03} yield quite different results
which also affects the theoretical result (see Tab.~\ref{tab}).

The comparison between experimental and theoretical values  reveals
that for the air--water interface the renormalized charges on the water side
seem to be sufficient to explain the observed repulsions. In this case
charge renormalization is essential because the straightforward application
of the linear theory (Eq.~(\ref{eq:hurd})) with the bare charge gives 
$\beta U \sim 10^7\times(R/d)^3$, which is orders of magnitude off.  
For the oil--water interface, Ref.~\cite{Ave02} reports tweezer measurement
data for the effective pair potential (PS spheres with sulfate groups, $R=1.35$
$\mu$m, $\sigma_c=8.9$ $\mu$C/cm$^2$) for two electrolyte concentrations
($\kappa^* \approx 2$ and 130).  
Both data sets could be fitted to one and the same pair potential.
Although the uncertainty in the contact angle translates into a considerable
spread of the theoretical predictions, the renormalized theory yields
a potential which is too small by at least a factor of 20.  
Therefore, the experimental results of Ref.~\cite{Ave02} point to still
another source of repulsion between the colloids.  In Ref.~\cite{Ave02} this
other source was argued to be colloidal surface charges on the oil side,
inferred only from the lack of a strong $\kappa$--dependence in the repulsion
as predicted by the linear theory.  This argument is insufficient since the
renormalized interaction weakens the $\kappa$--dependence considerably; the
hypothesis of possible extra charges on the oil side is rather supported by
the insufficient {\em magnitude} of the renormalized potential. 
Asymptotically the charges on the oil side together with their
image charges in the water create a net dipole in the nonpolar phase $p_{\rm
 oil}\approx 2q_{\rm oil}h$, where $h$ is their average distance from the
bottom of the particle.  
Due to the high dielectric constant of water, 
$p_{\rm oil}$  will be rather independent of the electrolyte concentration  
\cite{Dan06a}. 
The total effective dipole moment of the colloid is then given by 
$p_{\rm oil}+p_{\rm water}$ where $p_{\rm water}= 
(2\epsilon_1/\epsilon_2)q\kappa^{-1}g(\sigma^*,\kappa^*,\theta)$ 
is the dipole moment caused by the charges
on the water side.  The asymptotic interaction between two colloids is then
dominated by true dipole-dipole interactions given by
\bea
U(d)\approx 
\frac{1}{8\pi\,\epsilon_0\epsilon_1}{(p_{\rm oil}^2+2p_{\rm oil}p_{\rm water})}
\frac{1}{d^{3}}\;, 
\eea  
which act in addition to the interaction given in Eq.~(\ref{eq:hurd}).  
The results in Tab.~\ref{tab} suggest that $p_{\rm oil}$ is at least 
$\sqrt{20}$ times larger than $p_{\rm water}$ (for pure water). 
Even then, a certain electrolyte concentration dependence of the interaction
potential can be expected through the ensuing cross term $\propto p_{\rm
 oil}p_{\rm water}(\kappa^*)$ which has not been discussed in
Ref.~\cite{Ave02}. We note that recent, more extensive tweezer measurements at
an oil--water interface show indeed a marked dependence on the electrolyte
concentration \cite{Par07}.   

\begin{table}

 \begin{tabular}{ccllc}
   \hline \hline
   $\sigma_c^*$ & $\kappa^*$ &     
     \multicolumn{2}{c}{$\beta U/ (R/d)^3\times 10^3 $}  & $\theta$ \\ 
   &&{\em exp.}&\quad{\em theory}&(Ref.~\cite{Ave00}\dots
   \cite{Pau03}) \\ \hline
   \multicolumn{5}{c}{air/water -- Ref.\cite{Che06}}\\
    3900 & 1$\quad$& 8.06$\quad$& 1.8 \dots 4.6$\quad$& 30$^o$\dots 80$^o$\\
                 800 &1$\quad$ & 2.16 & 1.1 \dots 2.8 & 30$^o$\dots 80$^o$ \\ 
   \multicolumn{5}{c}{oil/water -- Ref.\cite{Ave02}}\\
    6800 & 2$\quad$ &  220 & 1.2 \dots 10 &
      75$^o$\dots 124$^o$ \\
             6800        & 135$\quad$ & 220 & 0.3 \dots 2.3 & 
      75$^o$\dots 124$^o$ \\ \hline \hline
 \end{tabular}
\caption{Comparison between available experimental data and
   Eq.~(\ref{eq:u_nonlin}) for the amplitude of the interaction potential. 
   For simplicity here $g_{\rm lin}=1$.} 
\label{tab}
\end{table}

In summary, within Poisson-Boltzmann theory we have discussed the
electrostatic interaction of charged spherical colloids trapped at an
interface between a nonpolar medium and water.  For charges on the water side
only, we have found a strong renormalization of the effective repulsion $U$,
changing the dependence on the surface charge density $\sigma_c$ and the
screening length $\kappa^{-1}$ from $U \propto \sigma_c^2 \kappa^{-2}$ (linear
theory) to 
$U \propto \ln^2[e\beta/(\epsilon_0\epsilon_2)\sigma_c\kappa^{-1}]$. For very
large charge densities, there is a possibility of a near independence of
the effective interactions on the salt concentration. Geometric effects
induced by the shape of the colloid are not expected to alter this result
significantly as long as $\kappa^{-1}$ is smaller than the linear size of the
colloid.  For colloids at an air--water interface, available experimental
results compare well with the renormalized theory, while for colloids at an
oil--water interface the renormalized theory underestimates the observed
effective potential, pointing to an additional source of repulsion such as
possible residual charges on the oil side.

{\em Acknowledgment:} M. O. acknowledges financial support from the DFG
through the Collaborative Research Centre ``Colloids in External Fields" SFB-TR6.

\pagebreak

\end{document}